\documentclass[conference]{IEEEtran}
\usepackage{graphicx} 

\usepackage{amsmath,amssymb,amsfonts}
\usepackage{graphicx}
\usepackage{textcomp}
\usepackage{hyperref}

\usepackage{booktabs}
\usepackage{multirow}
\usepackage{cite}
\usepackage{float}
\usepackage[table]{xcolor}
\usepackage{amsmath,amssymb}
\usepackage{algorithm}
\usepackage{algpseudocode}  
\usepackage{float}
\usepackage{xcolor}
\usepackage{colortbl}
\usepackage{array}
\definecolor{lightblue}{HTML}{D6EAF8}
\definecolor{headerblue}{HTML}{2E4057}

\definecolor{headerblue}{HTML}{2E4057}
\definecolor{row1}{HTML}{E8F4FD}
\definecolor{row2}{HTML}{F9F9F9}
\definecolor{accentgreen}{HTML}{4CAF50}

\algnewcommand\algorithmicinput{\textbf{Input:}}
\algnewcommand\INPUT{\item[\algorithmicinput]}
\algnewcommand\algorithmicoutput{\textbf{Output:}}
\algnewcommand\OUTPUT{\item[\algorithmicoutput]}
\begin{document}

\title{Beyond Black-Scholes: A Computational Framework for Option Pricing Using Heston, GARCH, and Jump Diffusion Models}

\author{
\IEEEauthorblockN{Karmanpartap Singh Sidhu, Pranshi Saxena}
\IEEEauthorblockA{Viterbi School of Engineering, University of Southern California\\
kssidhu@usc.edu, pranshis@usc.edu }
}

\maketitle

\begin{abstract}
This research tries to solve the issue of accurate option pricing in financial market, by employing the sophisticated models that go beyond the limitations of the traditional Black-Scholes model. Although, the Black-Scholes model provides a good closed -form solution for option pricing, but is limited by its presumptions like constant volatility, no dividend during the option's life, and inability to accommodate the sudden price movements or jumps in the asset prices. To overcome these limitations, this research uses Monte Carlo simulation alongside some advanced models like GARCH model, Heston model, Merton jump-diffusion model to help us get a better understand the dynamics of option pricing.

In this research \textbf{Monte Carlo simulation} serves as the backbone for the computational framework because of its ability to generate generation of thousands of simulated asset price paths with the help of geometric Brownian motion. All these models were implemented with the use of the Python programming language because of its simple syntax and its extensive libraries.

By combining the Black-Scholes model's simplicity and the flexibility of Monte Carlo simulations, the \textbf{Black-Scholes-Monte Carlo method} provides a dynamic approach to option pricing. In complex situations like path-dependent options or fluctuating market conditions, the classic Black-Scholes model is inadequate, even if it provides a closed-form solution for European-style options under assumptions like constant volatility and no dividends. By simulating diverse potential stock price pathways for the underlying asset using geometric Brownian motion, the Monte Carlo method overcomes these constraints. With Monte Carlo computational prowess, as we increase the number of simulations, greater becomes the accuracy, as opposed to traditional models which rely heavily on the assumptions.

\textbf{The GARCH (Generalized Autoregressive Conditional Heteroskedasticity) model} was employed to forecast the volatility of the underlying asset using the historical data for the specified strike price, capturing its time-varying nature and better resembling the real-life market conditions. Using the forecasted volatility from the GARCH model as input, simulated stock price paths were generated through Monte Carlo simulations. The simulated stock prices were then utilized to predict future option prices.

The \textbf{Heston-Monte Carlo model} uses the element of stochastic volatility along with the computational power of Monte Carlo simulations to price options more accurately. Unlike the Black-Scholes Monte Carlo model, which assumes constant volatility, the Heston model incorporates stochastic nature of the volatility using the historical data. By using this model, we are better able to capture real-life complex market dynamics market like volatility clustering and the volatility skew, which are often observed in the implied volatility of options.

In addition to the above models, this work incorporated the \textbf{Merton Jump Diffusion model, which} builds upon the Heston - Monte Carlo framework by incorporating the possibility of sudden price movements or jumps in the underlying asset. While the Black-Scholes Monte Carlo method assumes continuous price movements using the geometric Brownian motion, the Merton model adds a jump component, which is calculated using a Poisson process to simulate the random arrival of discrete price changes.

The results showed significant improvements in option pricing accuracy. The Black- Scholes Monte Carlo model provided better modelling with use of simulated stock path to improve the accuracy. The Heston model with use of Monte Carlo captured the stochastic nature of volatility, while the Heston jump diffusion model successfully incorporated the impact of sudden price jumps. The GARCH model provided better volatility forecasts, leading to more precise prediction of the future option prices. All these advanced models demonstrated their ability to address real-world market complexities. The results presented in this work were obtained from experiments conducted in November 2024, using live market data fetched at the time of execution.
\end{abstract}

\section{Introduction}

The traditional Black -- Scholes model has been used from around 1980s by providing the framework to determine the value of the option price (call -put). These models also helped the investors to make the hedging strategies to reduce the risk of the adverse price movement of the asset. This model was based on some key assumptions.

\begin{itemize}
    \item This is applicable only to the European style options, meaning the option exercised only at the expiration.
    \item The model assumes the constant volatility and risk- free rate over time.
    \item The model assumes the underlying asset doesn't pay dividend over time.
    \item It assumes that the market operates constantly and there are no sudden changes or jumps in the asset prices.
\end{itemize}

Despite Black- Scholes being widely the used model for its simplicity, the model find itself in problem due to the complex and dynamic nature of the market. Some of the problems this model suffers from and how this work tries to address them using advanced computational methods.

\begin{itemize}
    \item It assumes the volatility of the underlying asset is constant over the life of the option. In real market situation volatility fluctuates. During the 2008 financial crisis, volatility rose and out- of -- money put options became expensive. The Black -Scholes model underpriced these options due to constant volatility. To address the constant volatility problem, we used the \textbf{Heston model} to introduce the stochastic volatility, meaning the volatility is modeled as random process.

    \item The Black-Scholes model assumes the asset price follows the geometric motion, meaning the asset price is continuous, with no sudden jumps. But during the unexpected market events like the 2008 financial crisis there is sudden movement in the prices of the asset. To address this shortcoming the Black-Scholes model, we tried inculcating the \textbf{Merton -- jump diffusion model} in the Monte Carlo method to predict the price jumps. In addition to this, this research uses machine learning based optimizers to find the parameters for the Heston model.

    \item By using \textbf{the GARCH (Generalized Autoregressive Conditional Heteroskedasticity}) model to predict the future option prices as they better at recording the dynamic and time dependent nature of volatility has advantage over the traditional Black -Scholes model.
\end{itemize}

The results are improvement over the traditional Black-Scholes model. The Heston model consistently gave the estimates closer to the market option values across all the strike considered. The mean error between the Heston model and the market option price is smaller than the GBM (Black-Scholes model). However, the traditional GBM performed reasonably well when the strike price was lower than the current market price. Heston models slightly over- estimate the option price due to lack of historical data availability more than one year. When using the Merton jump- diffusion model to price options gives better results when aligned with the current market price, especially for the deep-in-money and at-the -money options, while the GBM model consistently underprices option across all strike prices. The GBM models seems to work well for the simple market conditions but lacks effectiveness due to constant volatility. The Merton jump diffusion model is good for volatile assets such as cryptocurrency, where there is sudden price jumps. The Heston model works best overall for capturing the prices of asset with dynamic volatility and options with long term expirations.

\section{Problem Statement}

The purpose for the research  is to develop a more accurate and efficient method for valuing the options. The traditional models, such as the Black-Scholes model, have been the cornerstone of the option pricing but come with several limitations. Some of the Black-Scholes limitations are Inability to handle varying volatility, Inaccurate for the American style option. Assume a constant risk -free rate. Assume that the volatility of the underlying asset is constant over time. Volatility is one the of the key components in option pricing, hardly stay constant in financial markets, it tends to fluctuate, cluster according to the market conditions. By leveraging new techniques, we can address these gaps and improve the robustness of the option pricing. One of the foremost goals is to develop accurate and efficient method, which can be used for pricing options by addressing the limitation of the traditional models like the Black-Scholes. Through this work, we are trying to achieve robustness, flexibility and practical application of option pricing model in the real-world complex environment. with usage of advanced computational method, we are trying to develop a model that captures the behavior of the underlying asset in the real market scenario, at the same time maintaining the computational efficiency. To do that this research employed various models like

\begin{itemize}
    \item First, this starts with Black -Scholes monte Carlo method, which involved using the geometric Brownian motion to simulate the thousands of stock paths to get more accurate option value, which wouldn't have been possible with traditional model.

    \item The second method employed in this work is the Heston model, which is a improvement over the traditional Black Scholes Monte Carlo method by inculcating the stochastic volatility into the modeling.

    \item In this work we also used the Merton jump-diffusion model to include the real condition like sudden movements of the asset prices due to various economic or geopolitical factors. This model is more valuable in providing the realistic framework for option pricing in complex markets.

    \item To check for the robustness and accuracy of the approaches used, in this work the usage of GARCH model was employed to predict the future prices of the option and match with the market prices to check the efficiency of our model by addressing the dynamic nature of volatility by modelling it as time dependent and conditional on historical returns.
\end{itemize}

The results from the models employed will be compared to the real-world prices to perform the sensitivity analysis of the performance of the model by varying the key parameters like the volatility, time to maturity and strike prices to see how the option price change over time. Not only this, but the work also tries to compare the various models used and see in which situation the model works well and in which situation model deviates from the market values. The goal of the work is not only to address the limitations of the Black-Scholes model but also help in providing a unified framework for option pricing which can adapt to the various market conditions. With all these improvements over the traditional methods can help provide for the accurate valuation of options for the traders, help financial managers prepare a better hedging strategies and investment decisions. This is trying to connect the existing gap between the theatrical assumptions and real-life market complexities, also this help in the advancement of financial modeling of the financial instruments but at the same time it empowers the market participants to maneuver the market complexities.

\section{Approach}

In this research, four different mathematical models: `Monte Carlo Method', `Monte Carlo Method with Merton Jump Diffusion Model', `Monte Carlo Method with GARCH' and Heston Model have been developed or estimating option pricing. Furthermore, Machine Learning framework has been introduced to optimize parameters for the Heston and Merton Jump Diffusion models. The selected Stocks to perform the simulations are: `Tesla', `Meta', `AMC Entertainment Holding', `MARA Holdings' and `Shopify INC.' The live and historical data has been picked by calling Yahoo Finance API. They have a high volume of trading options and have diverse business models. The description is given below:
\definecolor{lightblue}{HTML}{D6EAF8}
\definecolor{headerblue}{HTML}{2E4057}

\begin{table}[H]
\centering
\caption{Description of Stock Portfolio Selection}
\renewcommand{\arraystretch}{2}
\begin{tabular}{|>{\bfseries\centering\arraybackslash}m{1.5cm}|>{\centering\arraybackslash}m{5.5cm}|}
\hline
\rowcolor{headerblue}
\color{white}\textbf{Stock} & \color{white}\textbf{Description} \\
\hline
\rowcolor{lightblue}
TSLA & Electric automobile multinational manufacturer \\
\hline
META & Multinational social media and AI technology conglomerate \\
\hline
\rowcolor{lightblue}
AMC & Leading cinema theatre chain in North America \\
\hline
MARA & Digital asset technology company focused on mining cryptocurrencies \\
\hline
\rowcolor{lightblue}
SHOP & Canada-based e-commerce company \\
\hline
\end{tabular}
\end{table}

\subsection{Monte Carlo Method for Option Pricing}

Monte Carlo method was developed in the late 1940s, during World War 2 era and named after the Monte Carlo Casino in Monaco, since the name should be associated with randomness and probability. Initially developed for solving problems to neutron diffusion and later was adopted in various fields, including finance. ``Monte Carlo method for Option Pricing'' \cite{ref1} was first introduced in 1980s in the paper by P Boyle \cite{ref1}. ``The described method simulates the process generating the returns on the underlying asset and invokes the risk neutrality assumption to derive the value of the option'' \cite{ref1}.

Later, more advanced works like ``introducing antithetic variates and control variates to reduce variance in option pricing simulations'' \cite{ref2} and ``reducing number of simulations while maintaining the accuracy for path-dependent options'' \cite{ref3} were further developed.

In this work, the stock price path of an asset is simulated over time utilizing Geometric Brownian Motion, which is a continuous-stochastic process, assumes that the logarithmic of the stock follows a Brownian motion with drift.

The stochastic differential equation for Geometric Brownian Motion is defined as:

\begin{equation}
dS(t) = \mu S(t)\,dt + \sigma S(t)\,dW(t)
\end{equation}

Where $S(t)$ is the asset price modeled, $\mu$ is the drift rate, $\sigma$ is the volatility of the asset price, $dW(t)$ is the wiener process for adding randomness in the price movement.

This is further discretized to simulate asset price for discrete time intervals given by:

\begin{equation}
S_{t+\Delta t} = S_t \exp\left[\left(r - \frac{1}{2}\sigma^2\right)\Delta t + \sigma Z_t\right]
\end{equation}

where $r$ is the risk-free interest rate, $\Delta t$ is the time increment for each step, $Z_t$ is the random variable

\subsubsection{Implementation}

For this method, the goal is to compare the estimated option price with the current. Initial stock data, including Strike Price, Volatility, Option Types are picked from Yahoo API for every stock and their different contracts. Initial parameters for the simulation are defined as: Number of Simulation' -- 10000, Time Step' -- 100, Interest Rate'- 0.047, and Option Type'- Call. The main Monte Carlo function is implemented that simulates the asset price using the discretized GBM equation, where each path represents a possible future scenario for the stock price. At the option's expiration, the payoff for each simulated path is calculated. Here, the call option payoff has been calculated which is defined as:
\begin{equation}
\text{Payoff} = \max(S_T - K, 0)
\end{equation}
where $S_T$ is the simulated asset price at maturity, $K$ is the chosen strike price for an option contract
The expected payoff is discounted back to the present by utilizing the risk-free rate, defined as:
\begin{equation}
\text{Option Price} = e^{-rT} \times \text{Average Payoff}
\end{equation}
Here $e^{-rT}$ is the discount factor and $T$ is the expiration time. The mean of all discounted payoffs over all simulated paths is used to determine the final option price for a contract.
The algorithm for this implementation is provided below:
\begin{algorithm}[H]
\caption{Option Pricing with Monte Carlo Simulation}
\label{alg:mc}
\small
\setlength{\itemsep}{1pt}
\setlength{\parsep}{1pt}
\setlength{\topsep}{1pt}
\begin{algorithmic}
\INPUT Stock Data, Strike Prices, Simulations, Steps, Option Type, $r$
\OUTPUT Estimated Option Price, Market Option Price
\Statex
\Statex \textbf{Step 1 --- Fetch Live Data}
\Function{Get\_Latest\_Data}{ }
    \State Extract: Stock Price, IV, Expiration Date, Market Price
    \State Calculate: Expiration Time (years) per Strike Price
    \State \Return Contract Data for each Strike Price
\EndFunction
\Statex
\Statex \textbf{Step 2 --- Initialize Parameters}
\State $r \gets 0.0427$, \; $N_{\text{sim}} \gets 10\,000$, \; $N_{\text{steps}} \gets 100$, \; Type $\gets$ Call
\Statex
\Statex \textbf{Step 3 --- Monte Carlo Simulation}
\Function{Monte\_Carlo\_Method}{$S_0,\; dt$}
    \For{each simulation path}
        \For{$t = 1$ \textbf{to} $N_{\text{steps}}$}
            \State $Z \sim \mathcal{N}(0,1)$; update price via discretized GBM
        \EndFor
    \EndFor
    \State Payoff $\gets \max(S_T - K,\;0)$
    \State \Return $e^{-rT} \times \overline{\text{Payoff}}$, Simulated Paths
\EndFunction
\Statex
\Statex \textbf{Step 4 --- Evaluate}
\State Compare estimated option price with market price
\end{algorithmic}
\end{algorithm}

\subsubsection{Results}

Below are the results found for estimated market price, and current market price (which was November 2024 when the experiments were done) for different Strike values for Shopify INC and Mara Holdings. 
\textbf{SHOPIFY:}
\begin{table}[H]
\centering
\caption{Estimated Option Price Vs Market Price for Shopify}
\renewcommand{\arraystretch}{1.5}
\small
\begin{tabular}{|>{\bfseries\centering\arraybackslash}m{1.5cm}|>{\centering\arraybackslash}m{2.2cm}|>{\centering\arraybackslash}m{2.2cm}|}
\hline
\rowcolor{headerblue}
\color{white}\textbf{Strike} & \color{white}\textbf{Est.} & \color{white}\textbf{Market} \\
\hline
\rowcolor{lightblue}
80 & 24.60 & 24.65 \\
\hline
81 & 23.60 & 23.28 \\
\hline
\rowcolor{lightblue}
82 & 22.60 & 22.90 \\
\hline
83 & 21.60 & 22.40 \\
\hline
\rowcolor{lightblue}
84 & 20.60 & 20.62 \\
\hline
85 & 19.60 & 19.80 \\
\hline
\end{tabular}
\end{table}

\begin{figure}[H]
\centering
\includegraphics[width=0.9\columnwidth]{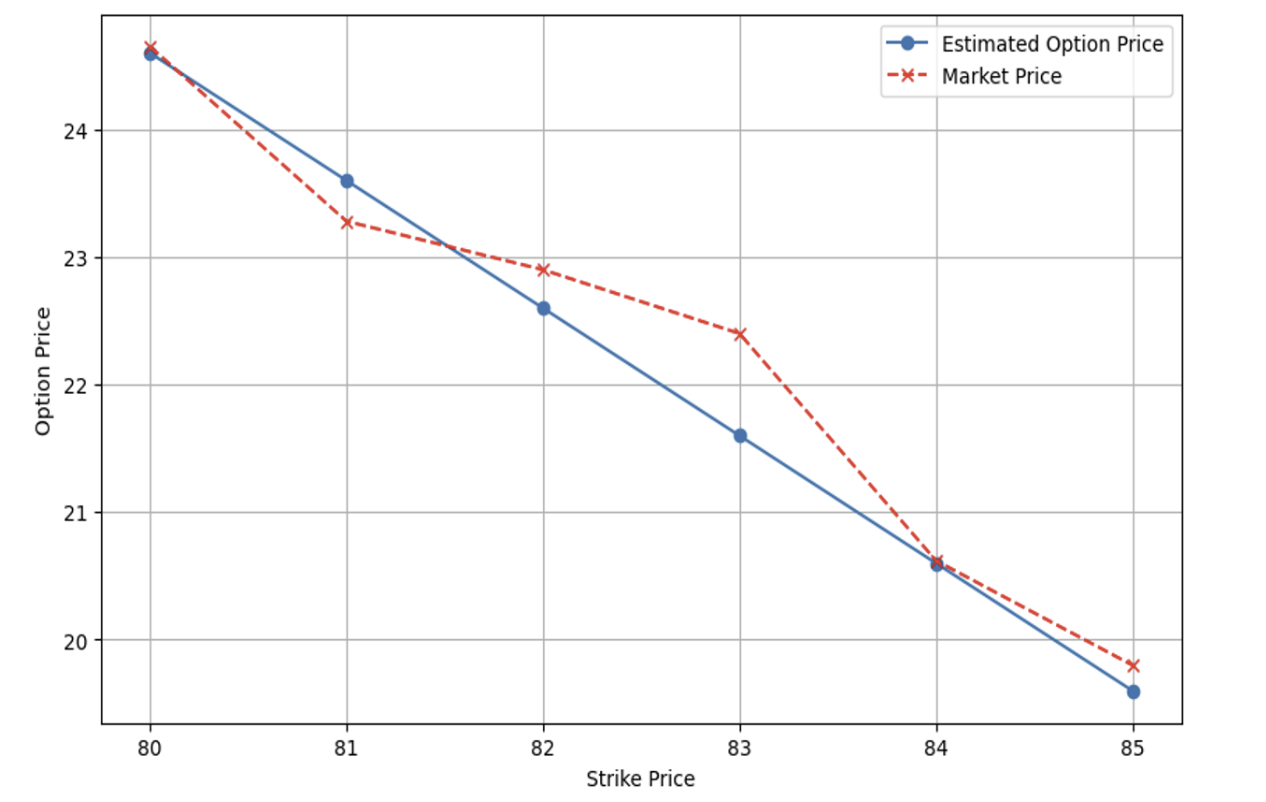}
\caption{Estimated Option price Vs Market price for SHOPIFY}
\end{figure}

General trend: market price doesn't show the monotonically decreasing trend with increasing the strike price. While the estimated prices follow more consistent downward trend as the strike price increases.

Deviation: Monte Carlo simulation is close to market prices at the lowest (\$80) and highest (\$84) strike prices. Significant deviations occur, especially at a strike price of (\$83), where the market price spikes much higher than the estimated price.

Reason: The black- Scholes monte Carlo method doesn't account for the sudden movements or jumps in the prices. Also, the models don't incorporate the market dynamics like the bid- ask spread, supply-demand imbalances.

\textbf{MARA:}
\begin{table}[H]
\centering
\caption{Estimated Option Price Vs Market Price for Mara}
\renewcommand{\arraystretch}{1.5}
\begin{tabular}{|>{\bfseries\centering\arraybackslash}m{2.2cm}|>{\centering\arraybackslash}m{2.2cm}|>{\centering\arraybackslash}m{2.2cm}|}
\hline
\rowcolor{headerblue}
\color{white}\textbf{Strike Price} & \color{white}\textbf{Estimated Price} & \color{white}\textbf{Market Price} \\
\hline
\rowcolor{lightblue}
18 & 2.63 & 2.77 \\
\hline
18.5 & 2.23 & 2.27 \\
\hline
\rowcolor{lightblue}
19 & 1.80 & 1.77 \\
\hline
19.5 & 1.36 & 1.27 \\
\hline
\rowcolor{lightblue}
20 & 0.98 & 0.77 \\
\hline
\end{tabular}
\end{table}

\begin{figure}[H]
\centering
\includegraphics[width=0.9\columnwidth]{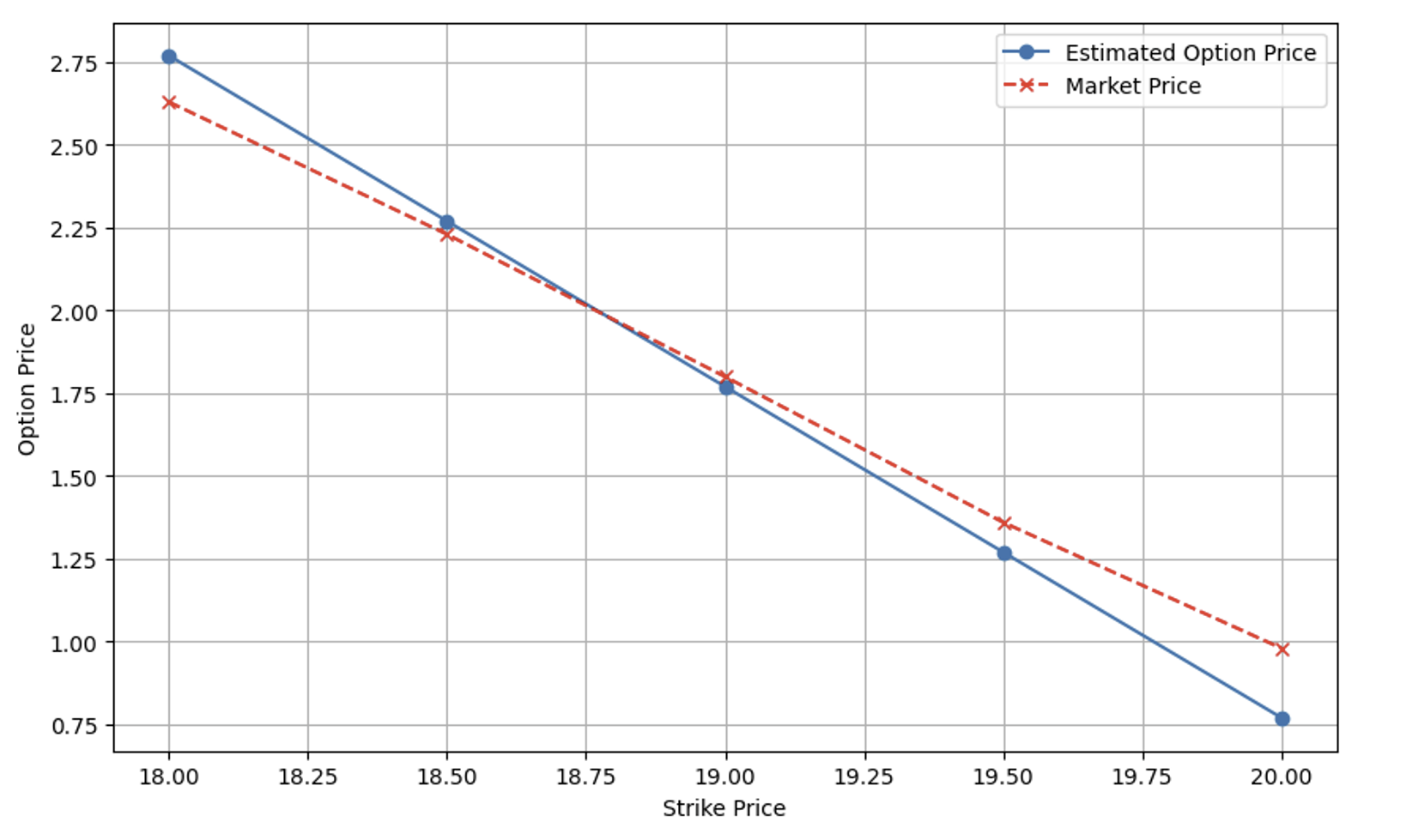}
\caption{Estimated Option price Vs Market price for MARA}
\end{figure}

General trend: Both market price and the estimated price decrease as the strike increase, which is conforms with theoretical expectation.

Deviation: Estimated price is little higher than market for lower strike and lower for the higher strike.

Reason: The model doesn't incorporate the factors like the transactional cost, supply -- demand imbalances. Since we are using the black. Scholes model with Monte Carlo simulation, the implied volatility once used at the start of simulation remains constant throughout the simulation of stock paths.

\subsection{Monte Carlo Method with Merton Jump Diffusion Model}

In this method, the Merton Jump Diffusion model is integrated into the original Monte Carlo method to further improve the accuracy of the estimation of option prices. This model was initially introduced \cite{ref4} to overcome one of the limitations of traditional methods in taking account for the large and sudden price changes, this issue was found with MARA stock, whose price fluctuation is high. The simulation of stock price can be re-written as:

\begin{equation}
dS(t) = S(t)\left[\mu\,dt + \sigma\,dW(t) + (J-1)\,dN(t)\right]
\end{equation}

where $J$ is the random Jump, $N(t)$ representing frequency of jumps

The jump component $(J-1)\,dN(t)$ represents for the jump of the stock price. $J$ is approximated as:

\begin{equation}
J = e^{Y},\ \text{where}\ Y \sim N\left(\mu_j, \sigma_j^2\right)
\end{equation}

The stock price path can be further calculated for small time interval $\Delta t$ as:

\begin{equation}
S_{t+\Delta t} = S(t)\exp\left[\left(r - \frac{\sigma^2}{2}\right)\Delta t + \sigma Z + \text{jumps}\right]
\end{equation}

where $r$ is the risk-free interest rate, $\Delta t$ is the time increment for each step, $Z_t$ is the random variable and

\begin{equation}
\text{jumps} = J \times \text{number of jumps}
\end{equation}

\subsubsection{Implementation}

In this method, the initial parameters and data for the stock price simulation is set similar to the prior method. And the final Option price is calculated in the similar fashion. For the extra introduced Jump component parameters, a new machine learning approach has been introduced to find the most optimized parameters. Different gradient-based optimizers have been tried to see which one minimizes the cost function. Based on these runs, L-BFGS-B optimizer is selected to estimate the values of jump component parameters ($\lambda_j, \mu_j, \sigma_j$).

The objective function is defined as the mean square difference between estimated option price and actual market price for selected strike prices of a Stock portfolio:

\begin{equation}
\text{Objective Function} = \sum\left(\text{Model Price}_i - \text{Market Price}_i\right)^2
\end{equation}

Initial guess values of jump component parameters are provided with correct bonds, and a Minimizer function is called to estimate the parameters, which are updated for every simulated path of the stock price. This approach provides us a way to analyze how well the Merton-Jump Diffusion model fits the observed market data. Since ($\lambda_j, \mu_j, \sigma_j$) have non-linear effect on the option price calculation, an optimizer-based approach is suitable to take account the non-linearity and explore parameter space to find the best fit. Moreover, Financial markets are driven by various factors, and calculating results based on a pure theoretical model can be difficult. Optimization based approach allows us for empirical calibration. Also, the L-BFGS-B method is particularly useful for problems with bound constraints, allowing to restrict parameters to realistic values. The detailed explanation of the optimizer is provided in the section 4.4.1.

The complete algorithm for the Merton Jump Diffusion Model is given below:

\begin{algorithm}[H]
\caption{Option Pricing via Merton Jump-Diffusion}
\label{alg:merton}
\footnotesize
\begin{algorithmic}
\INPUT Stock Data, Strikes, $r$, $\lambda_j$, $\mu_j$, $\sigma_j$
\OUTPUT Estimated \& Market Option Prices, Optimized Parameters
\Statex
\Statex \textbf{Step 1:} Fetch live data (same as Alg.~\ref{alg:mc})
\Statex \textbf{Step 2:} $r{=}0.0427$, $N_{\text{sim}}{=}10000$, $N_{\text{steps}}{=}100$, Call
\Statex
\Statex \textbf{Step 3 --- Optimize}
\State Init $\lambda_j{=}0.1$, $\mu_j{=}0.02$, $\sigma_j{=}0.05$
\State $(\lambda_j^*,\mu_j^*,\sigma_j^*) \gets \textsc{L-BFGS-B}\bigl(\sum_i(\text{Model}_i - \text{Mkt}_i)^2\bigr)$
\Statex
\Statex \textbf{Step 4 --- Simulate}
\For{each path, $t=1$ to $N_{\text{steps}}$}
    \State $Z{\sim}\mathcal{N}(0,1)$, $n_j{\sim}\text{Poi}(\lambda_j^* dt)$, $J_k{\sim}\mathcal{N}(\mu_j^*,\sigma_j^{*2})$
    \State Update price: GBM $+$ jumps
\EndFor
\State \Return $e^{-rT}\times\overline{\text{Payoff}}$
\Statex
\Statex \textbf{Step 5:} Compare estimated vs market prices
\end{algorithmic}
\end{algorithm}

\subsubsection{Results}

Below are the results found for estimated market price, and current market price for different Strike values for META and AMC Entertainment Holding. (All results are in Dollars)

\textbf{META:}

\begin{table}[H]
\centering
\caption{Estimated Option Price Vs Market Price for meta}
\renewcommand{\arraystretch}{1.5}
\begin{tabular}{|>{\bfseries\centering\arraybackslash}m{2.2cm}|>{\centering\arraybackslash}m{2.2cm}|>{\centering\arraybackslash}m{2.2cm}|}
\hline
\rowcolor{headerblue}
\color{white}\textbf{Strike Price} & \color{white}\textbf{Estimated Price} & \color{white}\textbf{Market Price} \\
\hline
\rowcolor{lightblue}
420 & 141.01 & 143.90 \\
\hline
450 & 110.97 & 106.27 \\
\hline
\rowcolor{lightblue}
470 & 91.52 & 84.39 \\
\hline
480 & 81.74 & 77.40 \\
\hline
\rowcolor{lightblue}
500 & 61.43 & 57.69 \\
\hline
\end{tabular}
\end{table}
\begin{figure}[H]
\centering
\includegraphics[width=0.9\columnwidth]{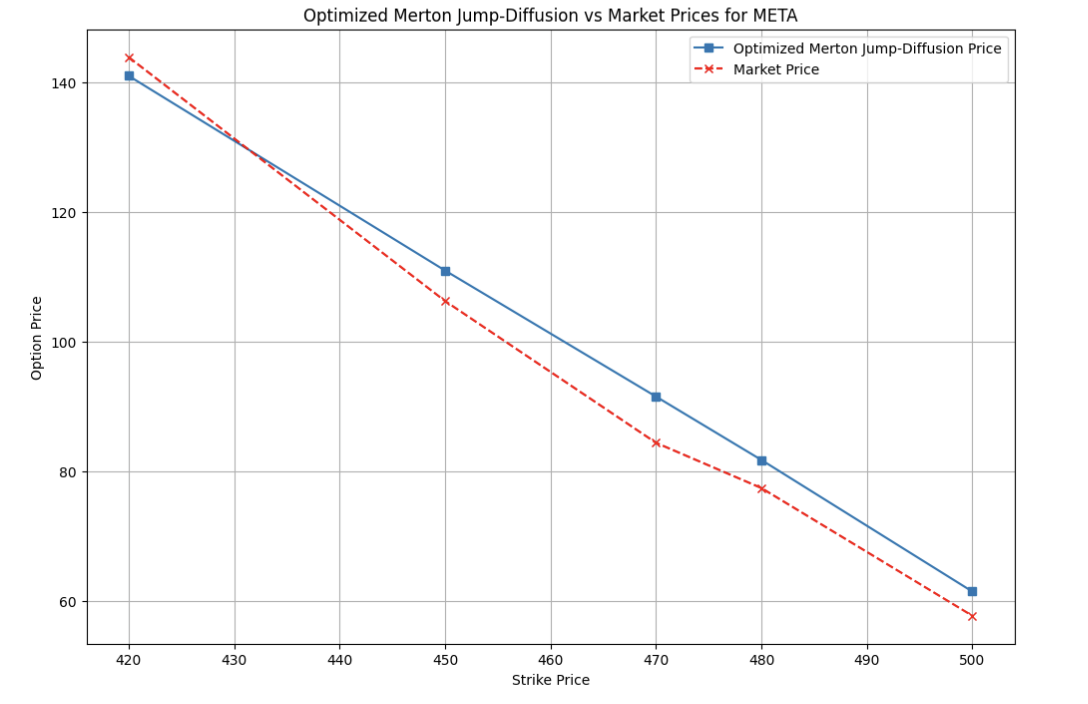}
\caption{Estimated Option price Vs Market price for META}
\end{figure}

General Trend: the market prices show a consistent decline as the strike price increases. The estimated price using the Merton jump -- diffusion model follows a similar downward trend but overestimate the market price at the higher strike price.

Deviation: the Merton jump -- diffusion model slightly underestimates the option price at the lowest strike price \$420 and for the prices \$450 and over it, the deviation increases as strike price rises.

Reason: the Merton jump -diffusion model incorporates jumps in the price of the underlying asset, which leads to higher option price estimates, especially at higher strike prices. Also, due to limited availability of the historical data, the stochastic nature of the volatility is not fully captured.

\textbf{AMC:}
\begin{table}[H]
\centering
\caption{Estimated Option Price Vs Market Price for amc}
\renewcommand{\arraystretch}{1.5}
\begin{tabular}{|>{\bfseries\centering\arraybackslash}m{2.2cm}|>{\centering\arraybackslash}m{2.2cm}|>{\centering\arraybackslash}m{2.2cm}|}
\hline
\rowcolor{headerblue}
\color{white}\textbf{Strike Price} & \color{white}\textbf{Estimated Price} & \color{white}\textbf{Market Price} \\
\hline
\rowcolor{lightblue}
3 & 1.31 & 1.18 \\
\hline
3.5 & 0.81 & 0.79 \\
\hline
\rowcolor{lightblue}
4 & 0.35 & 0.35 \\
\hline
4.5 & 0.05 & 0.08 \\
\hline
\rowcolor{lightblue}
5 & 0.01 & 0.04 \\
\hline
\end{tabular}
\end{table}
\begin{figure}[H]
\centering
\includegraphics[width=1\columnwidth]{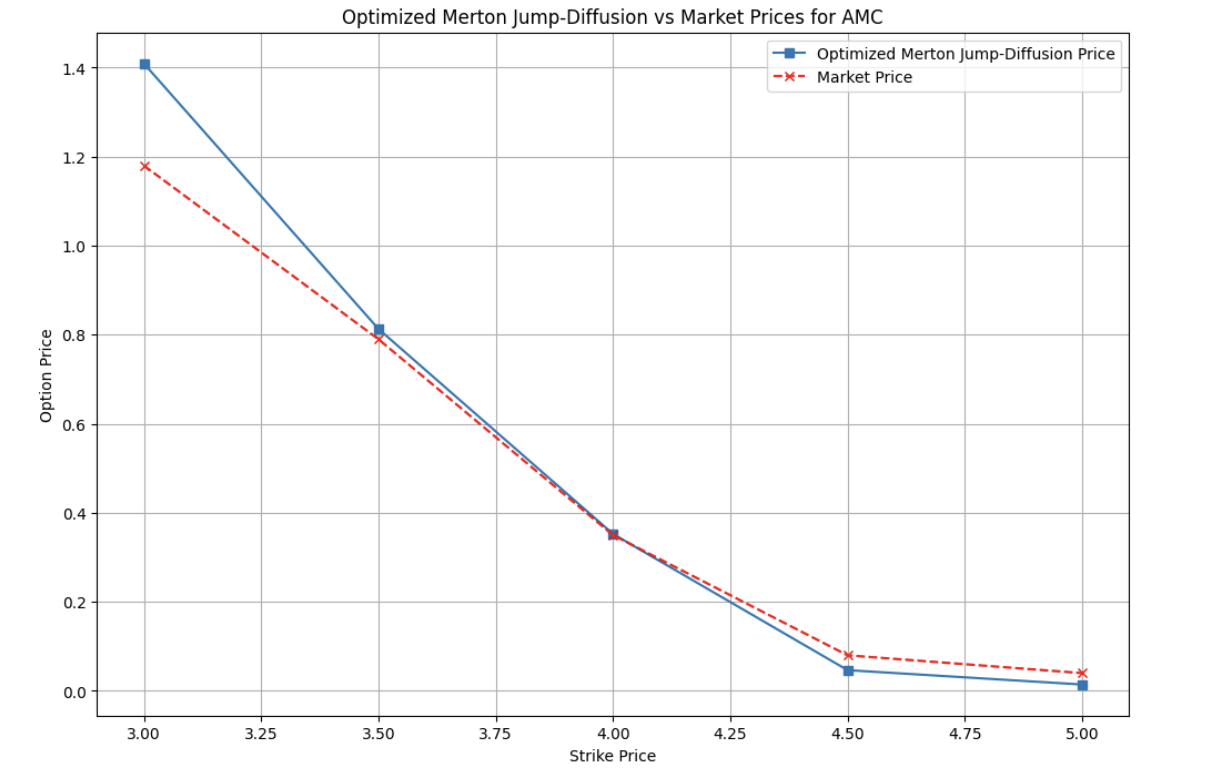}
\caption{Estimated Option price Vs Market price for AMC}
\end{figure}

General Trend: The market prices of the call options decline sharply as the strike price increases. Usually this is expected because higher strike prices reduce the intrinsic value with the likelihood of the option expiring in-the-money. The Merton jump- diffusion models follow the similar trend; however, the model slightly overestimates at lower strike values and vice -versa.

Deviation: The model tends to slightly overestimate at low strike prices and underestimate as the strike price increases and the option approaches out-of-the-money scenarios.

Reason: the Merton jump model assumes that the sudden price movements, so if the actual stock volatility is high but does not show sharp increase in price, model can overestimate the value at lower strike price.

\subsection{Monte Carlo Method with GARCH Model}

In the third method, after analyzing the results from the Monte Carlo Methods implemented above, the extension of the model has been done to introduce GARCH Model to estimate future option prices. The GARCH (Generalized Autoregressive Conditional Heteroskedasticity) model is implemented to calculate future volatility of the chosen strike price based on the historical data. The main assumption of the model is that the financial market is driven by volatility clustering where the high volatility is followed by more high volatility and low volatility is followed by more low volatility and can be calculated based on previous information. For this method GARCH (1,1) is implemented, where conditional variance $\sigma_t^2$ is defined as:

\begin{equation}
\sigma_t^2 = \alpha_0 + \alpha_i \epsilon_{t-i}^2 + \beta_j \sigma_{t-j}^2
\end{equation}

where $\alpha_0$ constant term to ensure the variance is positive, $\alpha_1$ to measure impact of past shock on current volatility, $\beta_1$ to measure persistence of past volatility, $\epsilon_t^2$ is the error term, assumed to be normally distributed with mean zero and variance $\sigma_t^2$

The initial Condition defined as:

\begin{equation}
\alpha_0 > 0,\quad \alpha_i \geq 0,\quad \text{and}\quad \beta_j \geq 0.
\end{equation}

\subsubsection{Implementation}

For this method, the goal is to estimate future option price for next three days. The live data from Yahoo finance is fetched to get latest stock price, strike price, contract expiration date. Along with that, historical data is also fetched to capture the volatility clustering. Based on this, GARCH (1,1) model is fitted, and volatility is calculated. This is done utilizing arch package. This predicted volatility is then annualized to be put as input for the Monte Carlo simulation. Initial parameters for the simulation are defined as: `Number of Simulation' -- 10000, `Time Step' -- 100, `Interest Rate'- 0.047, and `Option Type'- Call. The main Monte Carlo function is implemented that simulates the asset price using the discretized GBM equation, where each path represents a possible future scenario for the stock price. At the option's expiration, the payoff for each simulated path is calculated. By utilizing the risk-free rate, the expected payoff is discounted back to next three days for which we want to estimate the option price. `Call' option has been chosen to perform the simulation.

The call option payoff has been calculated which is defined as:

\begin{equation}
\text{Payoff} = \max(S_T - K, 0)
\end{equation}

where $S_T$ is the simulated asset price at maturity, $K$ is the chosen strike price for an option contract

The expected payoff is discounted back to the present by utilizing the risk-free rate, defined as:

\begin{equation}
\text{Option Price} = e^{-rT} \times \text{Average Payoff}
\end{equation}

Here $e^{-rT}$ is the discount factor and $T$ is the expiration time. The mean of all discounted payoffs over all simulated paths is used to determine the final option price for a contract.

The algorithm for Option Pricing is written below:

\begin{algorithm}[H]
\caption{Option Pricing via GARCH-Monte Carlo}
\label{alg:garch}
\footnotesize
\begin{algorithmic}
\INPUT Current \& Historical Stock Data, Strikes, $r$, Forecast Horizon
\OUTPUT Estimated option prices for future dates
\Statex
\Statex \textbf{Step 1:} Fetch live data: price, expiration, time to expiry
\Statex \textbf{Step 2:} Fetch historical variance over chosen period
\Statex
\Statex \textbf{Step 3 --- Fit GARCH(1,1)}
\State Fit model to variance series $\to$ estimate $\alpha_0, \alpha_1,  \beta_1$
\Statex
\Statex \textbf{Step 4 --- Forecast Volatility}
\State Forecast conditional variance $\to$ $\sigma_{\text{ann}} = \sigma_{\text{daily}} \times \sqrt{252}$
\Statex
\Statex \textbf{Step 5:} $r{=}0.0427$, $N_{\text{sim}}{=}100000$, $N_{\text{steps}}{=}100$, Call/Put
\Statex
\Statex \textbf{Step 6 --- Simulate}
\State Run GBM with GARCH-forecasted $\sigma_{\text{ann}}$ (as Alg.~\ref{alg:mc}, Step 3)
\State \Return Future option prices per strike
\end{algorithmic}
\end{algorithm}

\subsubsection{Results}

The predicted market price for 19th\textsuperscript{th}, 20th\textsuperscript{th} and 21st\textsuperscript{st} November, 2024 has been calculated for different stocks are provided below. Please note that 
all values are in Dollars 

\textbf{TESLA:}

\begin{table}[H]
\centering
\caption{Predicted Future Option Prices for tesla}
\renewcommand{\arraystretch}{1.5}
\begin{tabular}{|>{\bfseries\centering\arraybackslash}m{1.5cm}|>{\centering\arraybackslash}m{1.8cm}|>{\centering\arraybackslash}m{1.8cm}|>{\centering\arraybackslash}m{1.8cm}|}
\hline
\rowcolor{headerblue}
\color{white}\textbf{Strike Price} & \color{white}\textbf{19\textsuperscript{th} Nov} & \color{white}\textbf{20\textsuperscript{th} Nov} & \color{white}\textbf{21\textsuperscript{st} Nov} \\
\hline
\rowcolor{lightblue}
170 & 172.89 & 173.29 & 174.05 \\
\hline
175 & 168.05 & 168.23 & 169.28 \\
\hline
\rowcolor{lightblue}
195 & 147.76 & 148.64 & 150.23 \\
\hline
200 & 142.80 & 144.08 & 145.50 \\
\hline
\rowcolor{lightblue}
205 & 138.05 & 138.55 & 140.40 \\
\hline
\end{tabular}
\end{table}

\textbf{SHOPIFY:}

\begin{table}[H]
\centering
\caption{Predicted Future Option Prices for shopify}
\renewcommand{\arraystretch}{1.5}
\begin{tabular}{|>{\bfseries\centering\arraybackslash}m{1.5cm}|>{\centering\arraybackslash}m{1.8cm}|>{\centering\arraybackslash}m{1.8cm}|>{\centering\arraybackslash}m{1.8cm}|}
\hline
\rowcolor{headerblue}
\color{white}\textbf{Strike Price} & \color{white}\textbf{19\textsuperscript{th} Nov} & \color{white}\textbf{20\textsuperscript{th} Nov} & \color{white}\textbf{21\textsuperscript{st} Nov} \\
\hline
\rowcolor{lightblue}
80 & 28.39 & 29.97 & 31.57 \\
\hline
81 & 27.52 & 29.10 & 30.71 \\
\hline
\rowcolor{lightblue}
82 & 26.72 & 28.23 & 29.79 \\
\hline
83 & 25.75 & 27.45 & 29.21 \\
\hline
\rowcolor{lightblue}
84 & 24.80 & 26.53 & 28.37 \\
\hline
\end{tabular}
\end{table}

\textbf{META:}
\begin{table}[H]
\centering
\caption{Predicted Future Option Prices for meta}
\renewcommand{\arraystretch}{1.5}
\begin{tabular}{|>{\bfseries\centering\arraybackslash}m{1.5cm}|>{\centering\arraybackslash}m{1.8cm}|>{\centering\arraybackslash}m{1.8cm}|>{\centering\arraybackslash}m{1.8cm}|}
\hline
\rowcolor{headerblue}
\color{white}\textbf{Strike Price} & \color{white}\textbf{19\textsuperscript{th} Nov} & \color{white}\textbf{20\textsuperscript{th} Nov} & \color{white}\textbf{21\textsuperscript{st} Nov} \\
\hline
\rowcolor{lightblue}
420 & 137.97 & 146.42 & 153.57 \\
\hline
450 & 112.20 & 123.29 & 134.04 \\
\hline
\rowcolor{lightblue}
470 & 95.80 & 109.22 & 120.87 \\
\hline
480 & 88.68 & 103.37 & 115.07 \\
\hline
\rowcolor{lightblue}
500 & 74.95 & 91.95 & 104.10 \\
\hline
\end{tabular}
\end{table}

\textbf{MARA:}
\begin{table}[H]
\centering
\caption{Predicted Future Option Prices for mara}
\renewcommand{\arraystretch}{1.5}
\begin{tabular}{|>{\bfseries\centering\arraybackslash}m{1.5cm}|>{\centering\arraybackslash}m{1.8cm}|>{\centering\arraybackslash}m{1.8cm}|>{\centering\arraybackslash}m{1.8cm}|}
\hline
\rowcolor{headerblue}
\color{white}\textbf{Strike Price} & \color{white}\textbf{19\textsuperscript{th} Nov} & \color{white}\textbf{20\textsuperscript{th} Nov} & \color{white}\textbf{21\textsuperscript{st} Nov} \\
\hline
\rowcolor{lightblue}
18 & 3.35 & 3.84 & 4.27 \\
\hline
18.5 & 3.00 & 3.59 & 4.02 \\
\hline
\rowcolor{lightblue}
19 & 2.67 & 3.28 & 3.74 \\
\hline
19.5 & 2.39 & 3.07 & 3.56 \\
\hline
\rowcolor{lightblue}
20 & 2.09 & 2.73 & 3.24 \\
\hline
\end{tabular}
\end{table}
\subsection{Heston Model for Monte Carlo Simulation}

In this method, the limitations of Black-Scholes model are addressed, where constant volatility is assumed. It was introduced in 1993 \cite{ref5}. The model allows to correlate a stock price and its volatility, which helps to analyze the asymmetric behavior of option prices and help in better predicting the future prices.

Here, the Heston model describes the underlying stock price and its volatility as:

\begin{equation}
dS(t) = S(t)\left(\mu\,dt + \sqrt{v(t)}\,dW_1(t)\right)
\end{equation}

\begin{equation}
dv(t) = \kappa\left(\theta - v(t)\right)dt + \sigma_v \sqrt{v(t)}\,dW_2(t)
\end{equation}

where $S(t)$ is the stock price at time $t$, $\mu$ is the drift rate, $v(t)$ is the variance, $\theta$ is long-term mean variance, $\sigma_v$ is the volatility, $W_1$ and $W_2$ are two Brownian motions

This is further discretized to simulate asset price for discrete time intervals given by:

\begin{equation}
S_{t+\Delta t} = S_t \exp\left[\left(r - \frac{1}{2}\sigma^2\right)\Delta t + \sigma Z_1\right]
\end{equation}

where $r$ is the risk-free interest rate, $\Delta t$ is the time increment for each step, $Z_1$ is the random variable for the stock price path

\begin{equation}
v_{t+\Delta t} = v_t + \kappa\left(\theta - v_t\right)\Delta t + \sigma_v \sqrt{v_t}\,Z_2
\end{equation}

where $Z_2$ is the standard normal random variance of the variance path, defined as:

\begin{equation}
Z_2 = \rho Z_1 + \sqrt{1 - \rho^2}\,\widetilde{Z}
\end{equation}

where $\widetilde{Z}$ is an independent standard normal random variable

The algorithm for the Heston Model implemented is described below.
\begin{algorithm}[H]
\caption{Option Pricing via Heston Model}
\label{alg:heston}
\footnotesize
\begin{algorithmic}
\INPUT Current \& Historical Stock Data, Strikes, $r$, Heston Parameters
\OUTPUT Estimated Option Price, Market Option Price
\Statex
\Statex \textbf{Step 1:} Fetch live data: price, IV, time to expiry
\Statex \textbf{Step 2:} Fetch historical log-returns $\to$ approximate variance
\Statex
\Statex \textbf{Step 3 --- Optimize Heston Parameters}
\State Init: $\kappa{=}2.0$, $\theta{=}\overline{V}$, $\sigma_v{=}0.3$
\State Set $V_0 = \text{IV}^2$; predict variance via Heston params
\State $(\kappa^*,\theta^*,\sigma_v^*) \gets \textsc{Minimize}\bigl(\sum_i(V_{\text{pred},i} - V_{\text{hist},i})^2\bigr)$
\Statex
\Statex \textbf{Step 4:} $r{=}0.0427$, $N_{\text{sim}} \in \{100,1000,10000,100000\}$, $N_{\text{steps}}{=}1000$
\Statex
\Statex \textbf{Step 5 --- Heston Simulation}
\For{each path, $t=1$ to $N_{\text{steps}}$}
    \State $Z_1,Z_2 \sim \mathcal{N}(0,1)$; update price \& volatility via Heston
\EndFor
\State Payoff $\gets \max(S_T{-}K,0)$ or $\max(K{-}S_T,0)$
\State \Return $e^{-rT}\times\overline{\text{Payoff}}$
\Statex
\Statex \textbf{Step 6:} Compare estimated vs market prices
\end{algorithmic}
\end{algorithm}

\subsubsection{Implementation}

For this method, the goal is to compare the estimated option price with the current. Initial stock data, including Strike Price, Volatility, Option Types are picked from Yahoo API for every stock and their different contracts. Initial parameters for the simulation are defined as: `Number of Simulation' -- 10000, `Time Step' -- 100, `Interest Rate'- 0.047, and `Option Type'- Call. From the historical data, Logarithmic difference between consecutive closing prices of the stock has been extracted to calculate the variance of that time period. In the past, Research has been done around to integrate machine learning with Heston Model to estimate its parameters, the authors of \cite{ref5} has used ``deep learning-based approach for forecasting volatility in the Heston model framework. They use long short-term memory (LSTM) networks to model the time series of variances and improve option pricing accuracy''. This paper \cite{ref6} ``used neural networks to model the implied volatility surface derived from the Heston model. The authors show that machine learning can significantly reduce the complexity of calibrating the model to market data, making it more efficient while maintaining high accuracy''. In this paper, \cite{ref7} parameters of the model are optimized by ``by treating the calibration process as a sequential decision problem, the reinforcement learning agent learns optimal strategies for parameter estimation, improving the calibration speed and precision compared to conventional optimization techniques''. Inspired by these methods, Machine learning based optimization method is introduced to fine tune the parameters for the Heston model. Different gradient-based optimizers have been tried to see which one minimizes the cost function. Based on these runs, L-BFGS-B optimizer is selected to estimate the values of the parameters in Heston model.

The cost function is defined to quantify the difference between the model estimated variance and the historical variance. By minimizing the difference, the model parameters $\kappa$, $\theta$, $\sigma_v$ can be estimated to match the predicted option price actual market price.

The objective Loss function is defined as:

\begin{equation}
\text{Loss}(\kappa, \theta, \sigma_v) = \sum\left(\sigma_{\text{obs},t}^2 - v_t\right)^2
\end{equation}

where $N$ is the total number of time steps, $\sigma_{\text{obs},t}^2$ is the historic variance at time $t$, and $v_t$ is the variance predicted by the Heston model.

The optimizer used to find the parameters is L-BFGS-B (Limited-memory Broyden-Fletcher-Goldfarb-Shanno with Box constraints) best for optimizing differentiable functions, and in this work, it provided best results. It is a gradient based optimizer, calculates the first derivative of the objective function to fine tune parameters, and approximates second order derivative (Hessian matrix) of the objective function.

Let $f(x)$ be the loss function defined above, where $x = (\kappa, \theta, \sigma_v)$. The gradient $\nabla f(x)$ with respect to the $x$ is calculated. The parameters $x = (\kappa, \theta, \sigma_v)$ is updated as:

\begin{equation}
x_{k+1} = x_k - \alpha_k H_k^{-1} \nabla f(x_k)
\end{equation}

where $\alpha_k$ is learning rate, $H_k^{-1}$ is inverse Hessian matrix at iteration $k$

It should be noted that instead of storing the complete Hessian matrix which is a square matrix of second order partial derivative of a scalar-valued function. It tells about the curvature of the function, by calculating eigen values. If all the eigen values of the matrix are positive, the optimizer is stuck in local minima, if all the eigen values are negative, then the optimizer is stuck in local maxima, and if the matrix is indefinite, meaning there are mixed signs eigen values, the function has saddle. This helps to determine the direction and step size for the parameter updates.

In the main Monte Carlo function, both the paths of stock price and volatility are simulated over the time period of contract expiration. At the option's expiration, the payoff for each simulated path is calculated. By utilizing the risk-free rate, the expected payoff is discounted back to next three days for which we want to estimate the option price. `Call' option has been chosen to perform the simulation.

The expected payoff is discounted back to the present by utilizing the risk-free rate, defined as:

\begin{equation}
\text{Option Price} = e^{-rT} \times \text{Average Payoff}
\end{equation}

Here $e^{-rT}$ is the discount factor and $T$ is the expiration time. The mean of all discounted payoffs over all simulated paths is used to determine the final option price for a contract.

The algorithm for the Heston Model implemented is described below.

\subsubsection{Results}

Below are the results found for estimated market price, and current market price for different Strike values for TESLA and AMC Entertainment Holding. (All results are in Dollars)

\textbf{AMC:}

\begin{table}[H]
\centering
\caption{Estimated Option Price Vs Market Price for amc}
\renewcommand{\arraystretch}{1.5}
\begin{tabular}{|>{\bfseries\centering\arraybackslash}m{2.0cm}|>{\centering\arraybackslash}m{2.2cm}|>{\centering\arraybackslash}m{2.2cm}|}
\hline
\rowcolor{headerblue}
\color{white}\textbf{Strike Price} & \color{white}\textbf{Estimated Price} & \color{white}\textbf{Market Price} \\
\hline
\rowcolor{lightblue}
3 & 1.37 & 1.39 \\
\hline
3.5 & 0.87 & 0.88 \\
\hline
\rowcolor{lightblue}
4 & 0.39 & 0.38 \\
\hline
4.5 & 0.08 & 0.10 \\
\hline
\rowcolor{lightblue}
5 & 0.01 & 0.05 \\
\hline
\end{tabular}
\end{table}

\begin{figure}[H]
\centering
\includegraphics[width=0.9\columnwidth]{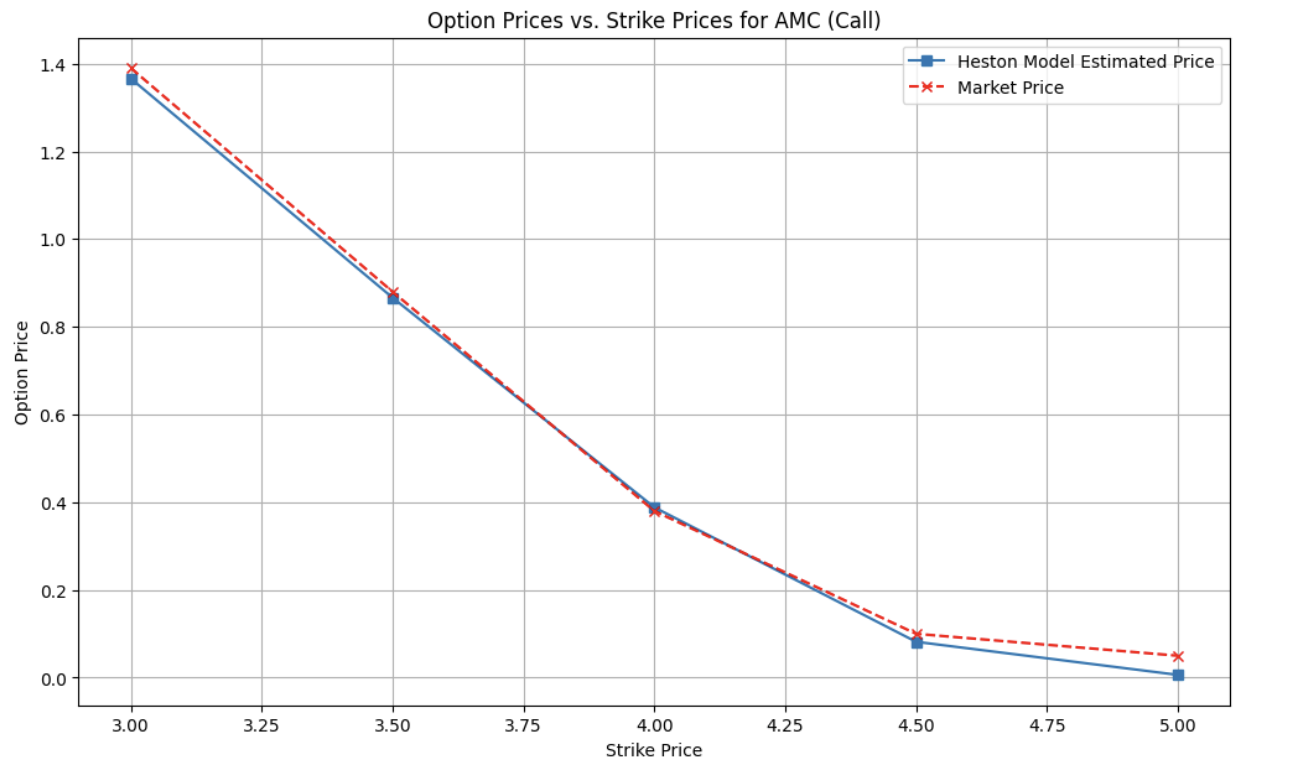}
\caption{Estimated Option price Vs Market price for AMC}
\end{figure}

General Trend: the market price and Heston model price decline as the value of the strike increases, which is in sync with theory.

Deviation: The deviation between the Heston model and the market price is almost minimal and is consistent, indicating the good relations between the model and the market data.

Reason: The slight mismatch can be due to market anomalies like liquidity issues, supply -demand imbalances, bid-ask spread

\textbf{TESLA:}
\begin{table}[H]
\centering
\caption{Estimated Option Price Vs Market Price for tesla}
\renewcommand{\arraystretch}{1.5}
\begin{tabular}{|>{\bfseries\centering\arraybackslash}m{2.2cm}|>{\centering\arraybackslash}m{2.2cm}|>{\centering\arraybackslash}m{2.2cm}|}
\hline
\rowcolor{headerblue}
\color{white}\textbf{Strike Price} & \color{white}\textbf{Estimated Price} & \color{white}\textbf{Market Price} \\
\hline
\rowcolor{lightblue}
170 & 151.37 & 149.53 \\
\hline
175 & 146.29 & 145.03 \\
\hline
\rowcolor{lightblue}
190 & 131.05 & 129.94 \\
\hline
195 & 126.88 & 125.25 \\
\hline
\rowcolor{lightblue}
200 & 120.89 & 120.96 \\
\hline
205 & 116.68 & 115.45 \\
\hline
\rowcolor{lightblue}
210 & 111.35 & 112.00 \\
\hline
215 & 105.96 & 106.61 \\
\hline
\rowcolor{lightblue}
220 & 101.10 & 100.11 \\
\hline
225 & 95.73 & 95.63 \\
\hline
\end{tabular}
\end{table}
\begin{figure}[H]
\centering
\includegraphics[width=0.9\columnwidth]{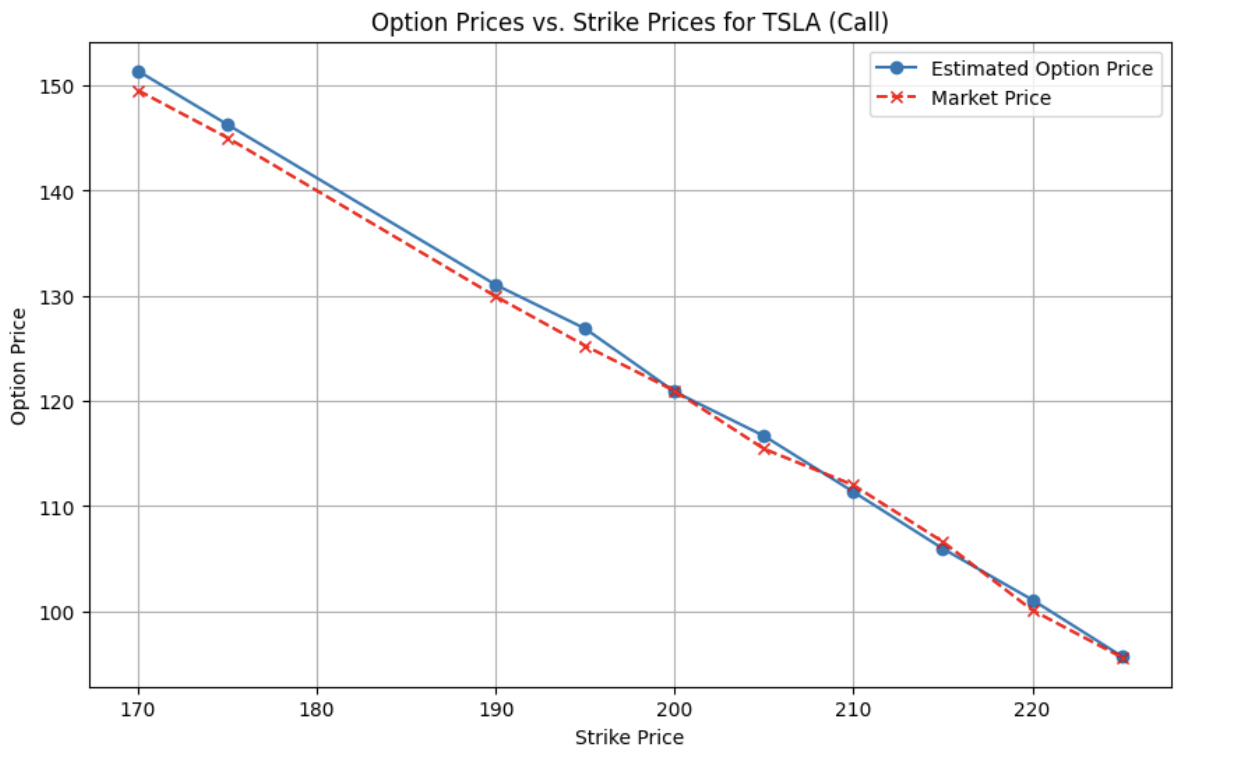}
\caption{Estimated Option price Vs Market price for TESLA}
\end{figure}

General Trend: The market price and Heston model price shows a consistent decline as the strike prices increases, which is in sync with theoretical expectation.

Deviation: for the most strike price the deviation is minimal. The most significant deviation occurs at the \$195, where the model overestimates the market price.

Reason: The large deviation at \$195 could be caused by some temporary market conditions, such as low liquidity or a large trade skewing the market price. Since the Heston model uses the historical data for the volatility and due to limited access to data, the volatility might not be accurately depicted in the model.

\section{Validation}

\subsection{Validation of Traditional Model vs Heston Model}

Here the results from the Heston model are validated with the traditional Monte Carlo method for Stock Tesla for a new run. This is being done to analyze how the Heston Model performs in comparison to Monte Carlo method, and it is validated, that it provides improvement in accuracy for estimating option price. (All values provided below are in dollars)

\begin{table}[H]
\centering
\caption{Estimated Option Price Vs Market Price for Monte Carlo Method and Heston Model for tesla}
\renewcommand{\arraystretch}{1.5}
\begin{tabular}{|>{\bfseries\centering\arraybackslash}m{1.2cm}|>{\centering\arraybackslash}m{1.6cm}|>{\centering\arraybackslash}m{1.6cm}|>{\centering\arraybackslash}m{1.6cm}|}
\hline
\rowcolor{headerblue}
\color{white}\textbf{Strike} & \color{white}\textbf{Heston} & \color{white}\textbf{Monte Carlo} & \color{white}\textbf{Market} \\
\hline
\rowcolor{lightblue}
170 & 150.35 & 151.69 & 149.53 \\
\hline
175 & 145.39 & 147.19 & 145.03 \\
\hline
\rowcolor{lightblue}
190 & 130.46 & 130.98 & 129.94 \\
\hline
195 & 125.41 & 126.40 & 125.25 \\
\hline
\rowcolor{lightblue}
200 & 120.42 & 120.68 & 120.96 \\
\hline
205 & 115.47 & 117.11 & 115.45 \\
\hline
\rowcolor{lightblue}
210 & 110.47 & 111.27 & 112.00 \\
\hline
215 & 105.44 & 106.03 & 106.61 \\
\hline
\rowcolor{lightblue}
220 & 100.44 & 101.28 & 100.11 \\
\hline
225 & 95.41 & 96.34 & 95.63 \\
\hline
\end{tabular}
\end{table}

\begin{figure}[H]
\centering
\includegraphics[width=0.9\columnwidth]{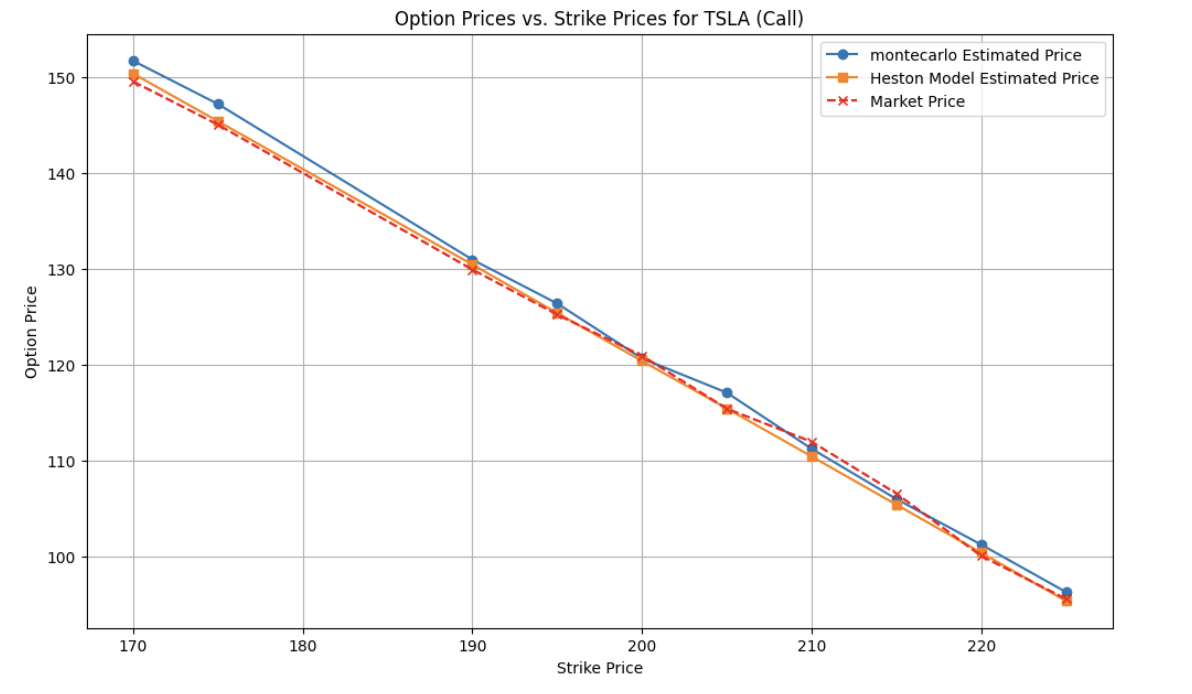}
\caption{ Estimated Option price Vs Market price for Monte Carlo Method and Heston Model for TESLA}
\end{figure}

General trend: the market prices and the estimated prices (Heston) show a smooth decline as the strikes price increases, whereas the estimated prices using the Black- Scholes (GBM) tend to slightly overestimate for the lower strike prices and slightly underestimate for higher strike prices.

Deviation: At \$170, the Heston model estimate is \$150.35 (closer to \$149.53 than GBM's \$151.69). Similarly, at \$225, the Heston estimate is \$95.41 (closer to \$95.63 than GBM's \$96.34). For lower strike prices (\$170--\$190), GBM slightly overestimates the option prices compared to the market. For higher strike prices (\$220--\$225), GBM slightly underestimates the market prices. The deviations range from a small overestimation at \$170 (\$151.69 vs. \$149.53) to a small underestimation at \$225 (\$96.34 vs. \$95.63).

Reason: GBM models assumes the constant volatility, which fails to incorporate the real market dynamics and at the same time the Heston models accommodate the real market scenario such as stochastic volatility, allowing it to better match the market prices. It was adjusted itself for the volatility skew which is often observed in the options.

\subsection{Validation of Predicted Future Prices by GARCH Model vs Actual Market Price}

Actual market price for November 19\textsuperscript{th} and 20\textsuperscript{th} for AMC Stock portfolio is compared with the predicted option price obtained from the GARCH model. (All values provided are in dollars). The results obtained from the GARCH model were ran previously to estimate option price for 19\textsuperscript{th} and 20\textsuperscript{th} November.

\begin{table}[H]
\centering
\caption{Estimated Prices Vs Market Price for Future Dates for amc}
\renewcommand{\arraystretch}{1.5}
\resizebox{\columnwidth}{!}{%
\begin{tabular}{|>{\bfseries\centering\arraybackslash}m{1.4cm}|>{\centering\arraybackslash}m{1.6cm}|>{\centering\arraybackslash}m{1.6cm}|>{\centering\arraybackslash}m{1.6cm}|>{\centering\arraybackslash}m{1.6cm}|}
\hline
\rowcolor{headerblue}
\color{white}\textbf{Strike} & \color{white}\textbf{Market (19\textsuperscript{th})} & \color{white}\textbf{Est. (19\textsuperscript{th})} & \color{white}\textbf{Market (20\textsuperscript{th})} & \color{white}\textbf{Est. (20\textsuperscript{th})} \\
\hline
\rowcolor{lightblue}
3 & 1.18 & 1.42 & 1.34 & 1.45 \\
\hline
3.5 & 0.79 & 0.86 & 0.93 & 1.02 \\
\hline
\rowcolor{lightblue}
4 & 0.35 & 0.32 & 0.42 & 0.46 \\
\hline
4.5 & 0.08 & 0.10 & 0.10 & 0.12 \\
\hline
\rowcolor{lightblue}
5 & 0.04 & 0.08 & 0.05 & 0.07 \\
\hline
\end{tabular}%
}
\end{table}

General Trend: the market price shows a decline in the value as the strike price increases, while the estimated prices using the GARCH model also depicts the similar pattern.

Deviation: The GARCH model predicted prices for some options, especially with the higher strike price are slightly higher than the current market prices. Also, the market price for the strike \$4.5 and \$5 drops more than predicted by the GARCH model.

Reason: There is some discrepancy between the forecasted volatility predicted by the model and actual volatility due to limited time access of the historical data for the specified strike price to train the model. With the model's dependency on the historical data to calculate the forecasted volatility, which may not capture the sudden jumps in the market sentiments. The relative slower decline at the strike price \$4.5 and \$5 can be attributed to the low liquidity for out -of -- money options in the market

\section{Conclusion}

In this work, a combination of advanced stochastic models and machine learning approaches has been used to solve the option pricing problem. Under normal assumptions, the Monte Carlo approach provided a solid foundation for pricing options, and its extension to include Merton jump-diffusion dynamics made it more realistic in capturing abrupt price fluctuations. The Heston model was used to further integrate stochastic volatility into the price prediction framework, giving it more depth. Machine learning-based optimizers were then used to fine-tune the parameters of the Merton jump-diffusion and Heston models, considerably increasing their accuracy and efficiency. The combination of these methodologies highlights the power of merging financial mathematics with data-driven approaches to solving complicated challenges in option pricing and risk management.

For the further extension of this work, neural networks can also be utilized to train and optimize model parameters, perhaps boosting the adaptability and precision of the estimates. Neural networks can learn subtle patterns in financial data, making them excellent for dynamic parameter optimization in stochastic models. In addition, quantum walks can also be examined as a novel way for simulating stock price and volatility routes. Quantum walks offer a probabilistic framework that may provide computing advantages and fresh insights into asset price movements. By merging these techniques, the research hopes to push the boundaries of classical and quantum computational approaches in financial engineering, providing unique tools for pricing complicated derivatives and controlling financial risks.

\section{References}

\bibliographystyle{IEEEtran}


\end{document}